\documentclass{aastex}           
\usepackage{graphicx,times}
\usepackage{natbib}
\usepackage{amssymb,amsmath}

\begin{document}

\title{Pulsation Analysis of the High Amplitude $\delta$ Scuti Star CW Serpentis}

\author{J. S. Niu}
\affil{Astronomy Department, Beijing Normal University,    Beijing 100875, China}
\author{J. N. Fu}
\affil{Astronomy Department, Beijing Normal University,    Beijing 100875, China}
\author{W. K. Zong}
\affil{Astronomy Department, Beijing Normal University,    Beijing 100875, China}

\begin{abstract}
Time-series photometric observations were made for the High Amplitude $\delta$ Scuti  star CW Ser between 2011 and 2012 at the Xinglong Station of National Astronomical Observatories of China.  After the frequency analysis of the light curves, we confirmed the fundamental frequency of $f = 5.28677\ cd^{-1}$, together with 8 harmonics of the fundamental frequency, 7 of which are newly detected. No additional frequencies were detected. The $O-C$ diagram, produced with the 21 newly determined times of maximum light combined with those provided in the literature, helps to obtain a new ephemeris formula of the times of maximum light with the pulsation period of $0.189150355 \pm 0.000000003\ days$.
\end{abstract}


\section{Introduction}
\label{sect:intro}

$\delta$ Scuti stars are a class of pulsating variable stars that lies in the classical instability strip cross the main sequence on the Hertzsprung-Russell diagram, whose pulsations are driven by the $\kappa-$mechanism which drives the pulsations of both Cepheids and the RR Lyrae stars as well. With masses between 1.5 and 2.5$M_{\odot}$, periods between 0.03 and 0.3 days, luminosities between 10 and 50 $L_{\odot}$, $\delta$ Scuti stars pulsate with amplitudes from mmag up to tenths of a magnitude \citep{Breger2000}.\\

The high-amplitude $\delta$ Scuti stars (hereafter HADS) are a subgroup of $\delta$ Scuti stars as slow rotators with one or two dominant radial modes whose amplitudes are larger than 0.1 mag, although some of them may  have low-amplitude, non-radial modes in addition to the main pulsation modes \citep{Poretti2003}. It is interesting to probe whether the non-radial mode is an overall property of the pulsations of all HADS.\\

Pulsations of the HADS CW Serpentis (hereafter CW Ser (= HIP 077798, $\alpha_{2000} = 15^{h}52^{m}10^{s}$, $\delta_{2000} = 06^{\circ}05'26''$, $ <V> = 11^{m}.83$, $P_{0} = 0^{d}.1892 $, $\Delta V = 0^{m}.47$, A-F) were discovered by \cite{Hoffmeister1935}. \cite{Gieren1975} made time-series photometry for CW Ser, derived two times of maximum light and the period value of $P = 0^{d}.18915054 \pm 0.00000001$. CW Ser has been observed only by a few runs since its pulsation period is rather long. The spectral type of CW Ser was classified as F by \cite{Halprin1983} and \cite{Lopez1990} by using the $uvby\beta$ photometry. \cite{Xu2002} calibrated it's spectral type as A-F. There were also several runs of photometric observations carried out in Konkoly observatory from 1975 to 2005 \citep{Gieren1975,Agerer1996,Agerer1998,Agerer1999,Agerer2001,Hubscher2005}.\\

In the following sections, we present a detailed study of the pulsations and period changes of CW Ser, mainly based on extended time-series photometric observations from 2011 to 2012 at the Xinglong Station of National Astronomical Observatories of China (NAOC). The organization of the paper is: Section 2 describes the data collection and reduction; Section 3 introduces the pulsation analysis of the new data; Section 4 discusses the calculated changes of the peirod with the $O-C$ method; Section 5 gives our conclusions.

\section{Observations and Data Reduction}
\label{sect:Obs}

Time-series photometric observations for CW Ser were made with the 85-cm telescope located at the Xinglong Station of NAOC between March 2011 and June 2012. The 85-cm telescope was equipped with a standard Johnson-Cousin-Bessel multicolour filter system and a PI1024 BFT CCD camera mounted on the primary focus \citep{Zhou2009}. The CCD camera has $1024\times1024$ pixels, corresponding to a field of view of $16.5^{'}\times16.5^{'}$, Since March 2012, this CCD camera was replaced by a PI512 BFT, which has $512\times512$ pixels corresponding to a field of view of $15^{'}\times15^{'}$. The observations were carried out through a standard Johnson $V$ filter with exposure time ranging from 20 to 150 seconds, depending on the atmospheric conditions. A journal of the new observations is listed in Table 1.

\begin{table}[!hrf]
\centering
\caption{Journal of the new observations in $V$ with the 85-cm telescope between 2011 and 2012.}
 \begin{tabular}{ccccc}
 \hline
CCD & Year & Month & Nights & Frames \\

 \hline
PI BFT1024 & 2011 & March & 4 & 1,249    \\
           &      & May & 7 & 2,247   \\
 \hline
PI BFT512  & 2012 & March & 5 & 830   \\
           &      & April & 2 & 245   \\
 \hline
 \end{tabular}
 \end{table}

In total, 4,571 CCD frames of data were collected for CW Ser within 18 nights. Figure 1  shows an image of CW Ser taken with the 85-cm telescope, where the comparison star and the check star are marked (see Table 2). After bias, dark and flat-field corrections, aperture photometry was performed by using the DAOPHOT program of IRAF. The light curves were then produced by computing the magnitude differences between CW Ser and the comparison star. The standard deviations of the magnitude differences between the check star and the comparison star yielded an estimation of photometry precisions, with the typical value of $0^{\rm m}.003$ in good observation conditions and $0^{\rm m}.011$ in poor cases. Although there were slight zero-point shifts, we adjusted it with the fit light curves of the data for every month (assuming the frequencies are stable in one month).

\begin{table}[!hr]
\centering
\caption{Information of the comparison and the check star.}
\begin{tabular}{ccccc}
\hline
ID&  Star-Name    &$\alpha$(2000)    &$\delta$(2000)    &$V$(mag)    \\
\hline
Comparison & USNOA2 0900-08292504  &$15^{h}53^{m}17^{s}.32$  &$+05^{\circ}59^{'}13^{''}.6$  &11.15  \\
Check & USNOA2 0900-08293690  &$15^{h}53^{m}24^{s}.81$  &$+05^{\circ}59^{'}15^{''}.5$  &13.50  \\
\hline
\end{tabular}
\end{table}

   \begin{figure}
   \centering
   \includegraphics[width=0.6\textwidth,height=0.6\textwidth]{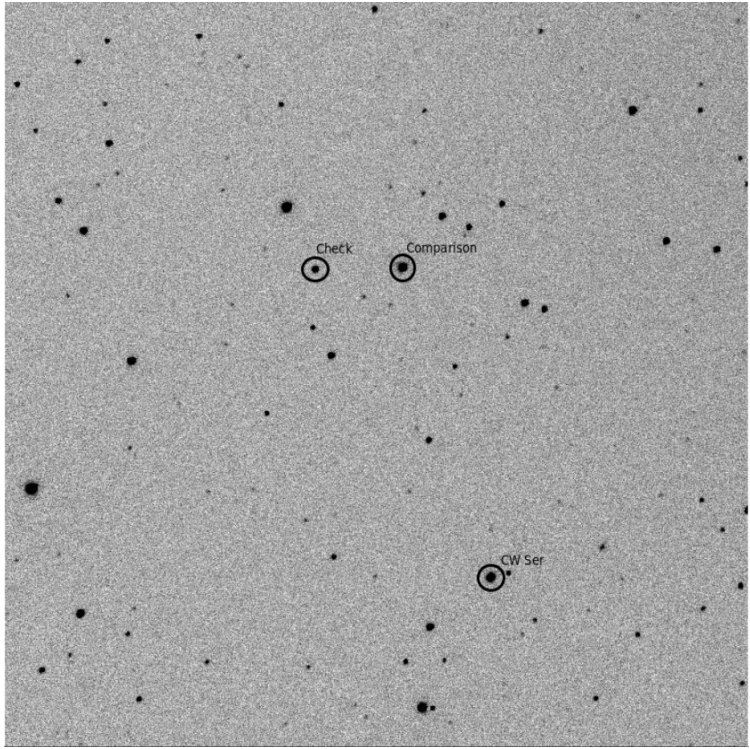}
   \caption{CW Ser, the comparison star and the check star, marked in an image taken with the 85-cm telescope. The field of view is $16.5'\times16.5'$. North is down and East is to the right.}
   \label{Fig1}
   \end{figure}

Figure 2 shows the light curves of CW Ser in Johnson \emph{V} band observed with the 85-cm telescope in 2011 and 2012, which were used to make frequency analysis.

   \begin{figure}
   \centering
   \includegraphics[width=\textwidth,height=0.5\textheight]{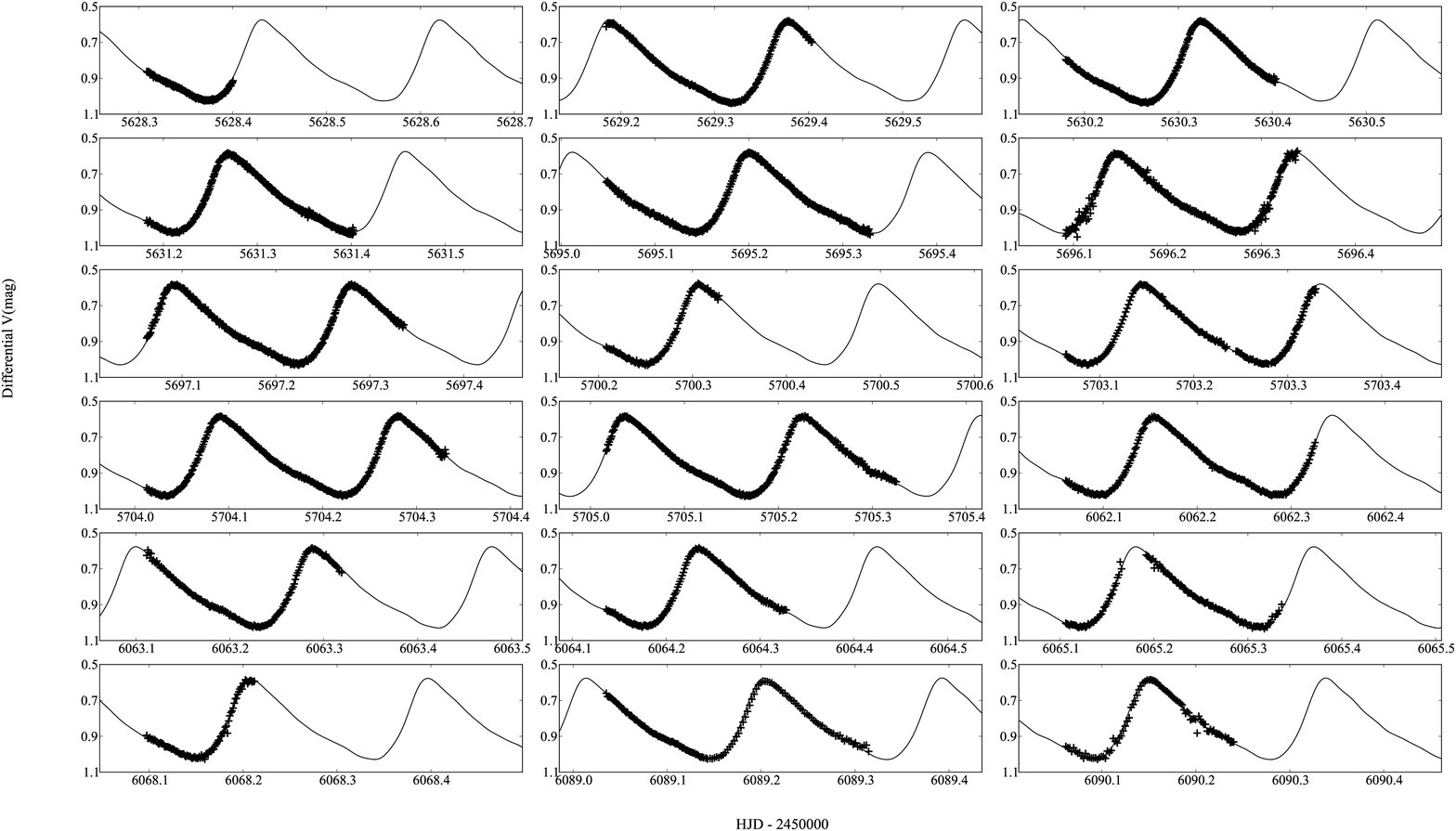}
   \caption{Light curves of CW Ser in the \emph{V} band from 2011 to 2012. The solid curves represent  the fitting with the 8-frequency solution listed in Table 3.}
   \label{Fig2}
   \end{figure}

\section{PULSATION ANALYSIS}
\label{sect:data}

Frequency analysis was performed with the light curves of CW Ser in 2011-2012 with the software PERIOD04 (\citealt{Lenz2005}), which makes Fourier transformations of the light curves to search for significant peaks in the amplitude spectra. The light curves are fitted with the following formula,
\begin{equation}
m=m_{0}+ \sum_{i} A_{i}\sin(2\pi(f_{i}t+\phi_{i})).
\end{equation}
The solutions with frequencies whose signal-to-noise ratios (S/N) are larger than 4.0 \citep{Breger1993} were listed in Table 3. The solid curves in Figure 2 show the fits with the frequency solution in 2011-2012. From Table 3, one notes that the 8 frequencies are composed of the fundamental frequency and  its 7 harmonics while no other frequencies in addition to the fundamental frequency and the harmonics are detected.

As can be seen from Figure 2, the constructed curves fit well the observed light curves, which shows that the fundamental frequency and its harmonics can explain the pulsation behavior of CW Ser.

Figure 3 shows the window function of the light curves in $V$ of CW Ser from 2011 to 2012. Figure 4 shows the amplitude spectra of the frequency pre-whitening process. Note that the peaks located in the low-frequency domain ($0-3\ c\ d^{-1}$) are not considered as significant signals of the variable star due to the instrument sensitivity instability and variations of sky transparency in the low frequency domain.

   \begin{figure}
   \centering
   \includegraphics[width=0.6\textwidth,height=0.4\textwidth]{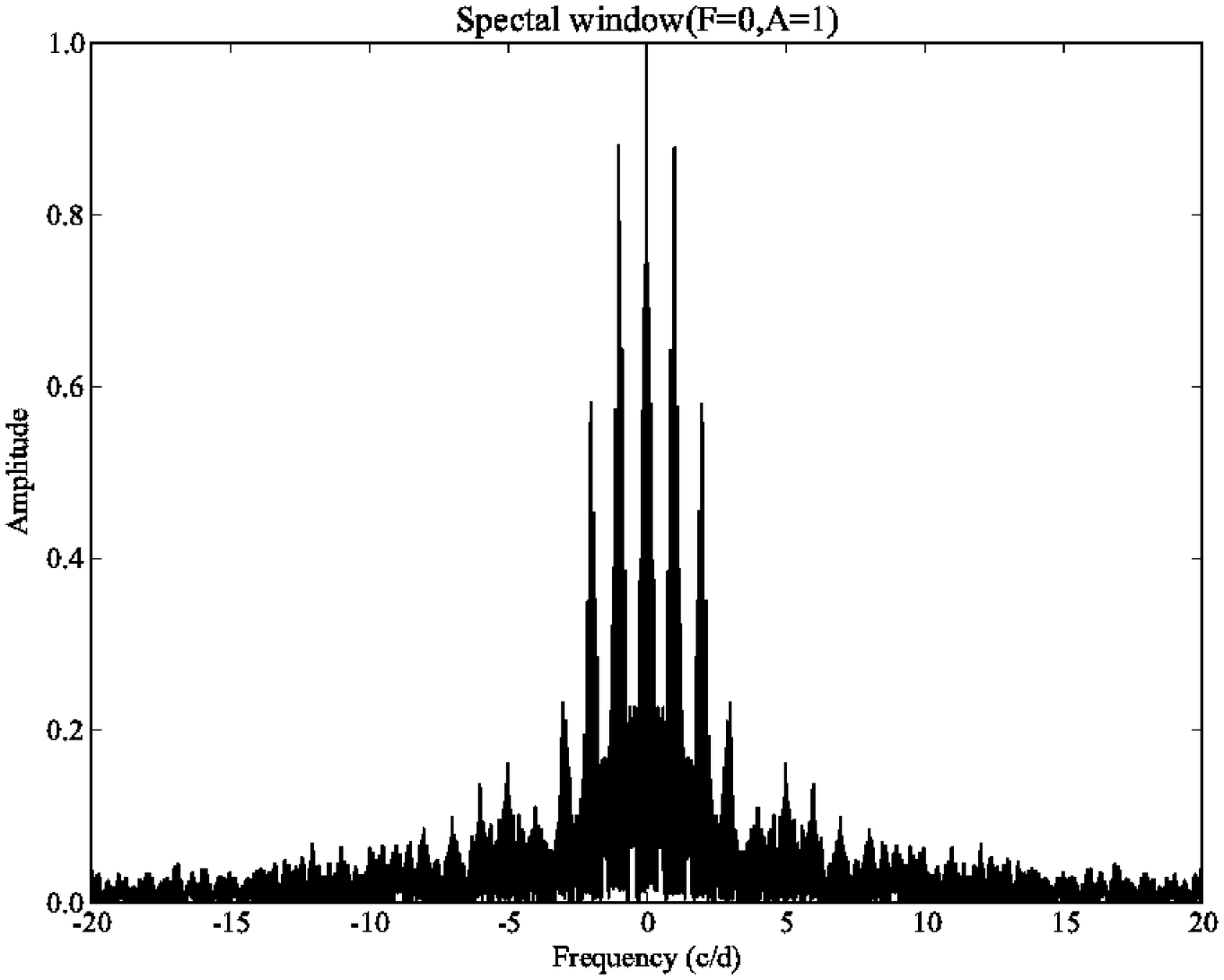}
   \caption{Spectral window of the light curves in $V$ for CW Ser from 2011 to 2012.}
   \label{Fig3}
   \end{figure}

\begin{table}[!hr]
 \centering
\caption{Multiple frequencies of CW Ser in 2011-2012.}
 \begin{tabular}{cccc}
 \hline
  × & Frequency ($c\ d^{-1}$) & Amp (mmg) & S/N \\
 \hline

$f_{0}$&        5.28677&         187.34&         47.1\\ 
$2f_{0}$&       10.57354&        66.37&         39.4\\ 
$3f_{0}$&       15.86032&        19.86&         27.2\\ 
$4f_{0}$&       21.14712&        10.11&         17.9\\ 
$5f_{0}$&       26.43660&        5.02&         15.6\\ 
$6f_{0}$&       31.72064&        2.83&         7.8\\ 
$7f_{0}$&       37.00211&        1.81&         6.0\\ 
$8f_{0}$&       43.29979&        1.04&         4.9\\ 
 \hline
 \end{tabular}

 \end{table}

   \begin{figure}
   \centering
   \includegraphics[width=\textwidth,height=0.6\textheight]{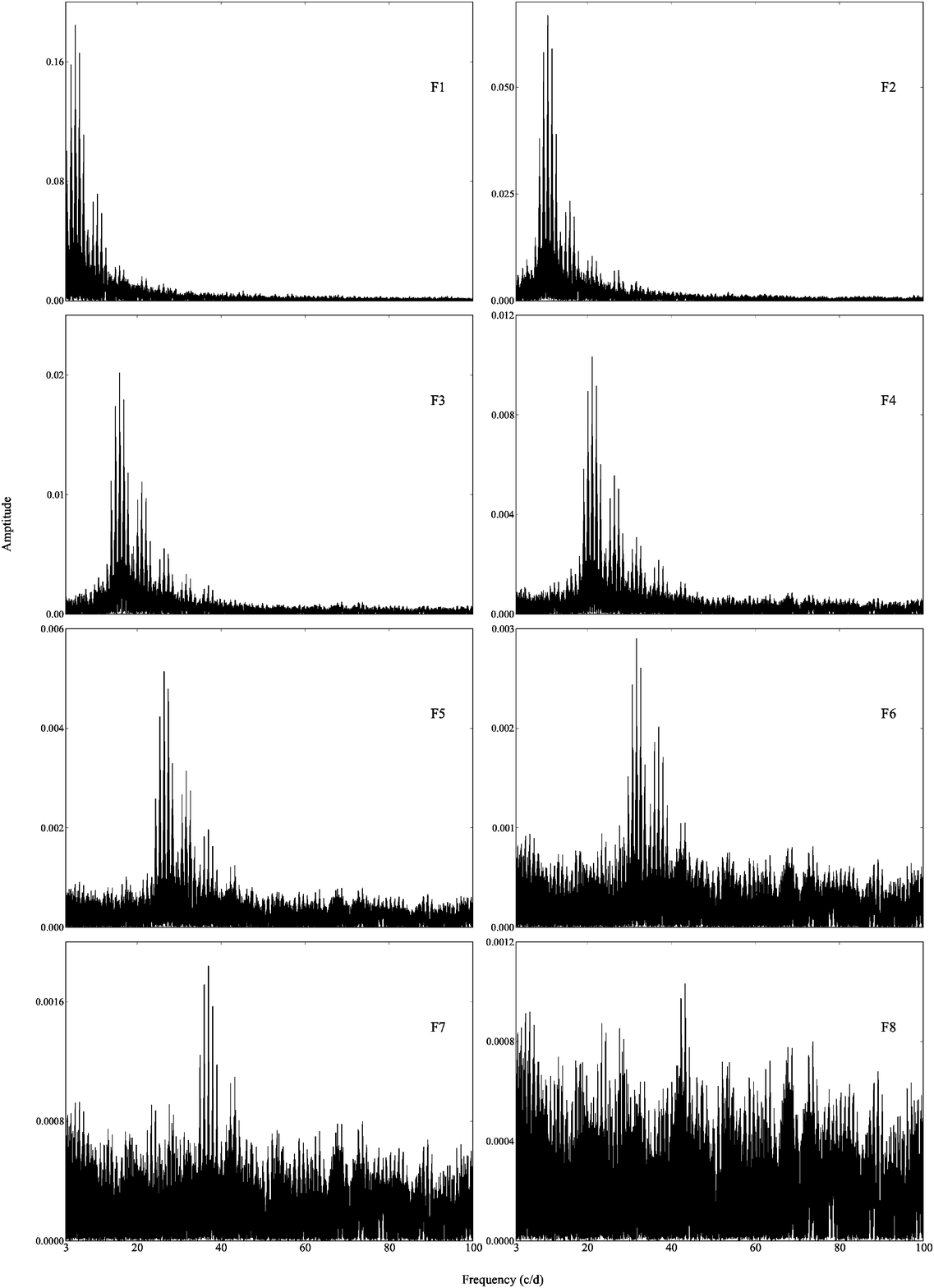}
   \caption{Fourier amplitude spectra of the frequency pre-whitening process for the light curves in V observed with the 85 cm telescope of NAOC in 2011-2012.}
   \label{Fig3}
   \end{figure}

\section{The $O-C$ Diagram}
With the new observations from 2011 to 2012, the light curves around the maximum light  were fitted with  third polynomials, which are sufficient to derive times of maximum light. 21 new times of maximum light in $V$ band are determined with the date in 2011-2012.

In order to make the $O-C$ analysis for the period change of CW Ser, the maximum times newly determined were combined with those provided by \cite{Gieren1975}, \cite{Agerer1996}, \cite{Agerer1998}, \cite{Agerer1999}, \cite{Agerer2001} and  \cite{Hubscher2005}. In total, 30 times of maximum light are collected and listed in Table \ref{tab8}, where the first time of maximum light is not included since we do not know the used detector and the data precision. In addition, there exists a large time gap between the first time of maximum light and the other data points.

The $O-C$ values are calculated and listed in Table 4 as well. The $O-C$ diagram is shown in Figure 5. 
A straight-line fit to the 30 times of light maxima yields the ephemeris formula,
\begin{equation}
  HJD_{max} = 2431212.3095(8) + 0.189150355(3) E
\end{equation}
, with a standard deviation of $\sigma = 0.0091(7)\ days$.\\

Since there are too free data points in the $O-C$ diagram, we do not try to make a second-order polynomial fit to estimate the period change rate.

   \begin{figure}
   \centering
   \includegraphics[width=0.6\textwidth, height=0.3\textwidth]{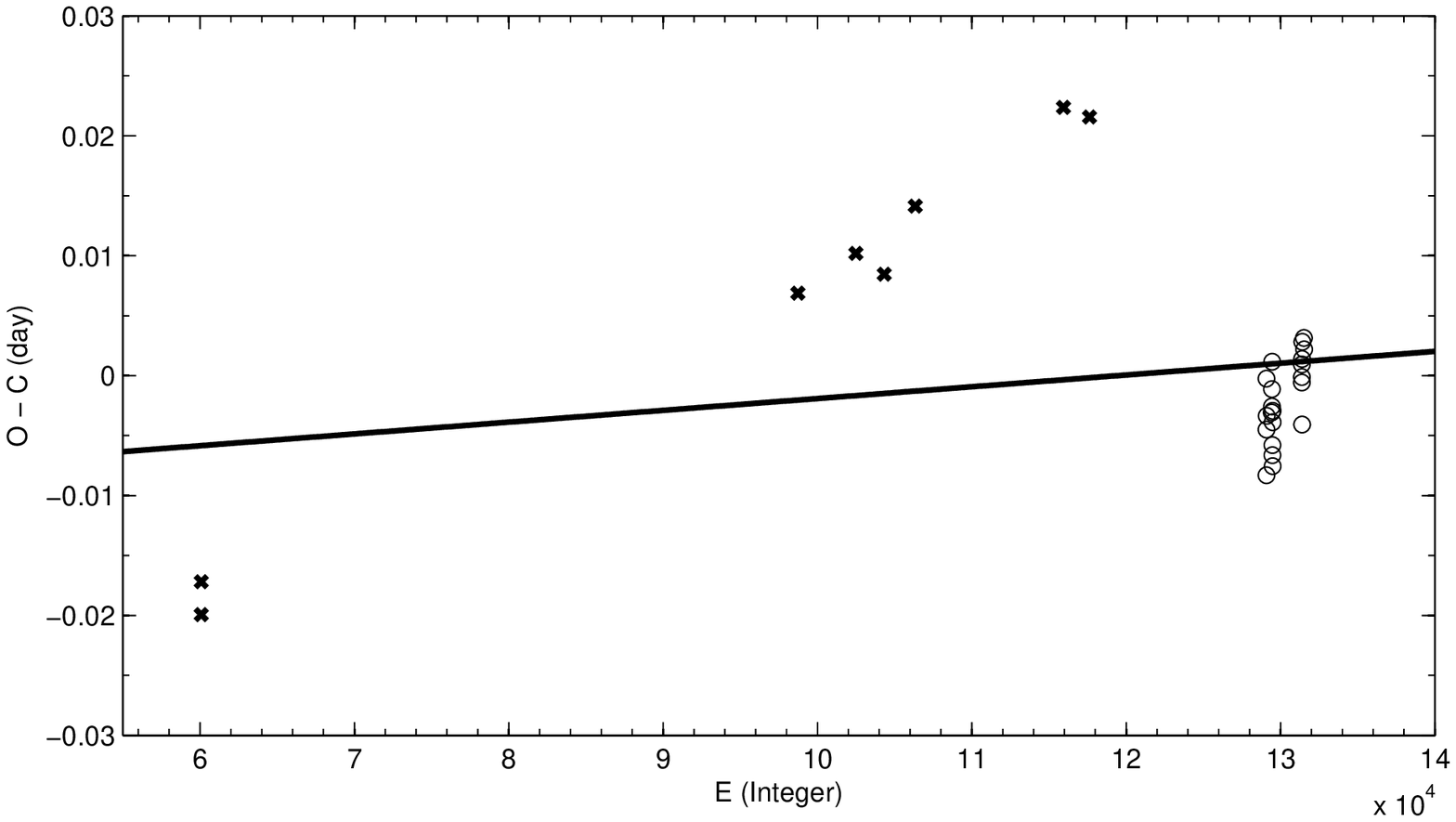}
   \caption{$O-C$ diagram constructed with the  data from the literature (cross) and the new data (open circle). The $O-C$ values are in days. E: cycle number. }
   \label{Fig4}
   \end{figure}

\section{Conclusions and Discussion}

On the basis of the time-series photometric data from 2011 to 2012, we determined  the fundamental frequency of pulsation as $5.28677 \pm 0.00001\ c\ d^{-1}$. No additional frequencies were resolved in the residual spectrum after prewhitening the fundamental frequency and its 7 harmonics. The light variations of CW Ser can be well reproduced with the fundamental radial mode and its 7 harmonics.\\

Analysis of the light curves of CW Ser suggests that no nonradial mode pulsations with amplitudes larger than $1.04\ mmag$ are detected, which shows that nonradial mode is not an overall property of the pulsations of all HADS at this amplitude level.

The 7 harmonics resolved in the pulsaitons of CW Ser indicate the distortion of the light curves from the standard sinusoidal function, which does not help the asteroseismological analysis for CW Ser since no new independent modes are presented. However, the existence of multiple harmonics shed lights on the nonlinear behaviour of the pulsation-excitation zone in the star.

According the $O-C$ method, we provide a new ephemeris formula  of the times of maximum light with the pulsation period of $0.189150355 \pm 0.000000003\ days$

More observations, includeing high duty-cycle time-series photometry and high quality spectroscopic observations, are extremely needed to caluculate the period change rate of CW Ser and deduce the properties of this HADS.\\


\section{Acknowledgments}

This work was supported by the National Natural Science Foundation of China (NSFC) under U1231202. The research is partially supported by the National Program of China (973 Program 2013CB834900) and the Fundamental Research Funds for Central Universities.


\clearpage

\begin{deluxetable}{ccccccc}
\tabletypesize{\footnotesize}
\tablewidth{0pc}
\tablecaption{Times ($T_{max}$ in HJD 2400000+ days) of maximum light of CW Ser. $E$: cycle number. O-C: in days. Det: Detector. pe: photoelectric photometer. Points not used in the O-C analysis are marked with an asterisk. S: Source.}
\tablehead{
\colhead{NO.} & \colhead{$T_{max}$} & \colhead{\emph{E}} & \colhead{$O-C$} & \colhead{Det.} & \colhead{Weight} &  \colhead{S}}

\startdata

       0&    31212.2800&        0&        ---&      ?&   ---&   (1)*\\
       1&    42575.5000&    60075&    -0.0171&      pe&  1.0&  (2)\\
       2&    42576.4430&    60080&    -0.0199&      pe&  1.0&  (2)\\
       3&    49888.4551&    98737&     0.0068&      CCD& 2.0&   (3)\\
       4&    50599.4746&   102496&     0.0102&      CCD& 2.0&   (4)\\
       5&    50944.4831&   104320&     0.0084&      CCD& 2.0&   (5)\\
       6&    51325.4376&   106334&     0.0141&      CCD& 2.0&   (6)\\
       7&    53143.3699&   115945&     0.0223&      pe&  1.0&  (7)\\
       8&    53462.6549&   117633&     0.0215&      pe&  1.0&  (7)\\
       9&    55628.3966&   129083&    -0.0082&      CCD& 2.0&   (8)\\
      10&    55629.3504&   129088&    -0.0002&      CCD& 2.0&   (8)\\
      11&    55630.2919&   129093&    -0.0045&      CCD& 2.0&   (8)\\
      12&    55631.2388&   129098&    -0.0033&      CCD& 2.0&   (8)\\
      13&    55695.1719&   129436&    -0.0030&      CCD& 2.0&   (8)\\
      14&    55696.3073&   129442&    -0.0025&      CCD& 2.0&   (8)\\
      15&    55697.2545&   129447&    -0.0011&      CCD& 2.0&   (8)\\
      16&    55700.2832&   129463&     0.0011&      CCD& 2.0&   (8)\\
      17&    55703.1135&   129478&    -0.0057&      CCD& 2.0&   (8)\\
      18&    55703.3018&   129479&    -0.0066&      CCD& 2.0&   (8)\\
      19&    55704.0575&   129483&    -0.0075&      CCD& 2.0&   (8)\\
      20&    55704.2503&   129484&    -0.0038&      CCD& 2.0&   (8)\\
      21&    55705.1970&   129489&    -0.0029&      CCD& 2.0&   (8)\\
      22&    56062.1276&   131376&     0.0009&      CCD& 2.0&   (8)\\
      23&    56063.2610&   131382&    -0.0005&      CCD& 2.0&   (8)\\
      24&    56064.2072&   131387&    -0.0001&      CCD& 2.0&   (8)\\
      25&    56065.1490&   131392&    -0.0040&      CCD& 2.0&   (8)\\
      26&    56065.3436&   131393&     0.0013&      CCD& 2.0&   (8)\\
      27&    56068.1823&   131408&     0.0028&      CCD& 2.0&   (8)\\
      28&    56089.1783&   131519&     0.0031&      CCD& 2.0&   (8)\\
      29&    56090.1231&   131524&     0.0021&      CCD& 2.0&   (8)\\

\enddata
\tablecomments{Sources: (1) \citet{Hoffmeister1935}; (2) \citet{Gieren1975}; (3) \citet{Agerer1996}; (4) \citet{Agerer1998}; (5) \citet{Agerer1999}; (6) \citet{Agerer2001}; (7) \citet{Hubscher2005}; (8) this work.}
\label{tab8}
\end{deluxetable}

\clearpage

\bibliographystyle{raa}
\bibliography{cwser}

\end{document}